# The neglected quantum effect on $CuO_2$ plane

Dong Hang

Institute of Applied Physics and Computational Mathematics

[Abstract]

The neglected quantum effect On $CuO_2$ plane--the transition of electron between neighbored ions is studied. To study this affection, an ideal square lattice composed of triangle unit is proposed and studied. Every unit in the square lattice has three ions and only one ion charge, the ion charge is distributed on one A ion and two B ions with different possibility, and the A-B bond length is far less than the distance between nearest units. The position difference between A-B ions results in a lattice deformation with fixed amplitude, and can be treated as an electricity dipole lattice. It is proved that such lattice has a negative energy gap that can be much larger than that of normal metal superconductor, and suggests a new superconductor mechanism.

Keywords: square lattice, triangle units, electronic dipole lattice, energy gap

It is well known that the superconductivity mainly occurs on the $CuO_2$ plane for cuprate superconductor [1,2], and some researches suggest that Cooper pairs may occur in real space [3], existing purely magnetic or electronic pairing mechanism [4], and so on. It is proved that the hole occurs on Cu-O ions with different possibility for various cuprate superconductors [5].

Though much study on this subject has been done, the effect of one kind of quantum phenomenon is neglected. On $CuO_2$ plane, there exists two kinds of hole-carriers, and it means that the electronegative of Cu and O ion is similar, so the transition of electrons between Cu and O ions will occur. On the other hand, since the ions are identical, to keep the occupying possibility of hole on every ion being same, the electron transition between same kinds of ions will occur. The transition of electrons between ions will violate the ion charge on $CuO_2$ plane, and it determines the character of $CuO_2$ plane.

The effort has been done to find the relationship between the character of $CuO_2$ plane and the paring mechanism [6]. As we know, the nature of electric conduction is the motion of electrons, therefore, on the $CuO_2$ plane, except for the $Cu^{2+}$, $Cu^{3+}$, $O^{2-}$, $O^{1-}$ ions, there must exist free electrons. For the free electrons, $Cu^{2+}$ and $O^{2-}$ can be treated as electrically neutral, they do not attract electron anymore; on the contrary, $Cu^{3+}$ and $O^{1-}$ can receive electrons and can be seen as cations, they form the ion lattice, and the total number of $Cu^{3+}$ and $O^{1-}$ should be approximately equal to the number of free electrons. From the viewpoint of statistics, if considering the repulsion between cations, every $CuO_2$ unit should has only one ion charge, and the charged $CuO_2$ unit should be approximately averagely distributed. The length of Cu-O bond is far less than the distance between charged $CuO_2$ units due to the low density of carriers.

In this paper, considering about the uncertainty caused by the complexity of electronic environment on $CuO_2$ plane, as a new idea attempting to explain some characteristic of cuprate superconductor, an ideal square lattice with triangle units is proposed and studied. The ideal square ion lattice is shown in Fig1, it is composed of triangle unit, every unit has two B ions and one A ion, the length of A-B bond is c, and is normalized by the side length of square. The situation that bond length being far less than distance between units is studied, which means $c \ll 1$.

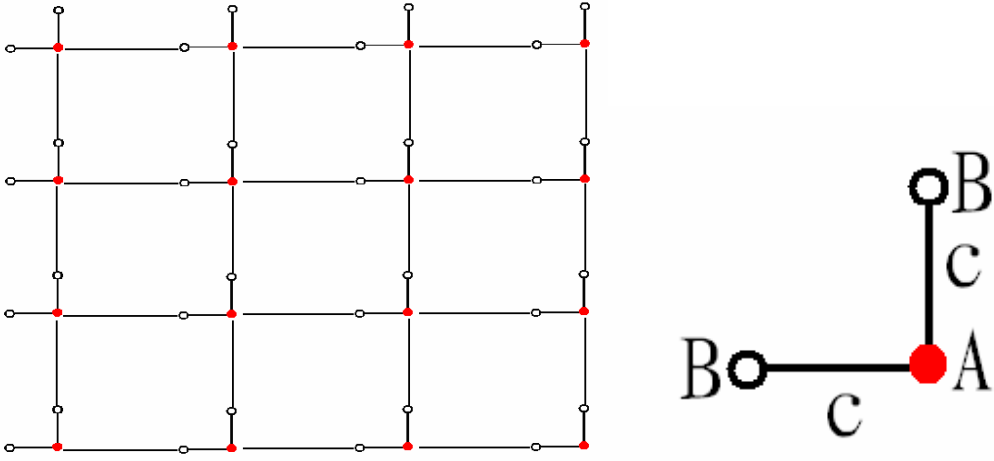

**Fig1. Illustration of triangle lattice**

The position difference between A and B ions will slightly violate the average ion charge distribution. Since c is far less than 1, the violation can be treated as a perturbation. When the charge transfers between A and B ions, it acts like another kind of lattice deformation, and one can expect that it results attractive interaction between electrons like normal metal superconductor.

Defining the occupancy probability on A ions and B ions as $P_A$ and $P_B$ respectively, then $P_A + 2P_B = 1$, means that one unit has only one charged ion. For simplicity, the coulomb screening is considered as $\varepsilon = \varepsilon_{eff}\varepsilon_0$, where $\varepsilon_0$ is electric permittivity, and $\varepsilon_{eff}$ stands for coulomb screening. The lattice energy comes from the interaction between one ion unit and the whole ion lattice can be written as:

$$E = \frac{e^2}{4\pi\varepsilon} \sum_{i,j} \left\{ \begin{array}{l} \frac{P_A^2}{\sqrt{i^2+j^2}} + \frac{2P_B^2}{\sqrt{i^2+j^2}} + P_B^2 \left[ \frac{1}{\sqrt{(i+c)^2+(j-c)^2}} + \frac{1}{\sqrt{(i-c)^2+(j+c)^2}} \right] \\ + P_A P_B \left[ \frac{1}{\sqrt{(i+c)^2+j^2}} + \frac{1}{\sqrt{(i-c)^2+j^2}} + \frac{1}{\sqrt{(j+c)^2+i^2}} + \frac{1}{\sqrt{(j-c)^2+i^2}} \right] \end{array} \right\} \quad (1)$$

By keeping the low order terms up to $c^2$ term, we have

$$\frac{1}{\sqrt{(i+c)^2 + j^2}} = \frac{1}{\sqrt{r_{ij}^2 * (1 + \frac{2ic + c^2}{r_{ij}^2})}} = \frac{1}{r_{ij}}\left(1 - \frac{1}{2}\frac{2ic+c^2}{r_{ij}^2} + \frac{3}{8}\left(\frac{2ic+c^2}{r_{ij}^2}\right)^2\right)$$

$$= \frac{1}{r_{ij}}\left(1 - \frac{1}{2}\frac{2ic+c^2}{r_{ij}^2} + \frac{3}{2}\frac{i^2 c^2}{r_{ij}^4}\right) \qquad r_{ij} = \sqrt{i^2 + j^2} \tag{2}$$

And the lattice energy is calculated as:

$$E = \frac{e^2}{4\pi\varepsilon}\sum_{i,j}\left\{\frac{P_A^2}{r_{ij}} + \frac{2P_B^2}{r_{ij}} + \frac{P_B^2}{r_{ij}}\left[2 + \frac{c^2}{r_{ij}^2}\right] + \frac{P_A P_B}{r_{ij}}\left[4 + \frac{c^2}{r_{ij}^2}\right]\right\}$$

$$= \frac{e^2}{4\pi\varepsilon}(P_A + 2P_B)^2 \sum_{i,j}\frac{1}{r_{ij}} + \frac{e^2}{4\pi\varepsilon_0}(P_B^2 + P_A P_B)\sum_{i,j}\frac{c^2}{r_{ij}^3}$$

$$= \frac{e^2}{4\pi\varepsilon}\sum_{i,j}\frac{1}{r_{ij}} + \frac{e^2}{4\pi\varepsilon}\frac{1-P_A^2}{4}\sum_{i,j}\frac{c^2}{r_{ij}^3} = E_0 + E_{dipole} \tag{3}$$

Where $e$ is charge of electron, and the $i$ and $j$ represent the position of different units on the lattice. One can see that $E_{dipole}$ is far less than $E_0$, this means that the approximation what we used, that is just keeping the low order term up to $c^2$ term, is reasonable. On the other hand, one can find that a negative c, standing for opposite direction, does not affect the characteristic of the square lattice.

The first term $E_0$ comes from the regular ion lattice, means that all ion charge being on A ions, and it keeps electricity equilibrium with free electrons; the interaction between free electrons and regular ion lattice is the same as that of normal metal, and it forms the ground state of the square lattice model.

The second term $E_{dipole}$ comes from the position difference between A-B ions, is the innate character of two-dimensional square lattice with triangle lattice. In the case of an ion charge transfers from A to B, it violates the charge equilibrium slightly. Since the $E_{dipole}$ acts like the interaction between electronic dipoles, here we named it after electronic dipole lattice, and the ion lattice can be seen as combination of regular lattice and dipole lattice. The interaction between electrons and electronic dipole lattice will induce additional energy, and then dominate the characteristic of the square lattice model.

From equation (3) one can find that the concentration of electronic dipoles $N_{dipole}$ is as follows:

$$N_{dipole} = (1-P_A^2)^{1/2}/2 \qquad (4)$$

The creation of electronic dipole lattice can be explained by means of electricity image method. Considering that c is small, the electricity image method, what widely used in classical electrodynamics, can be directly applied here: when the charge transfer to B from A, it acts like an negative ion charge creating on A site, and then to form an electronic dipole with positive ion charge on B, finally these electronic dipoles form the electronic dipole lattice.

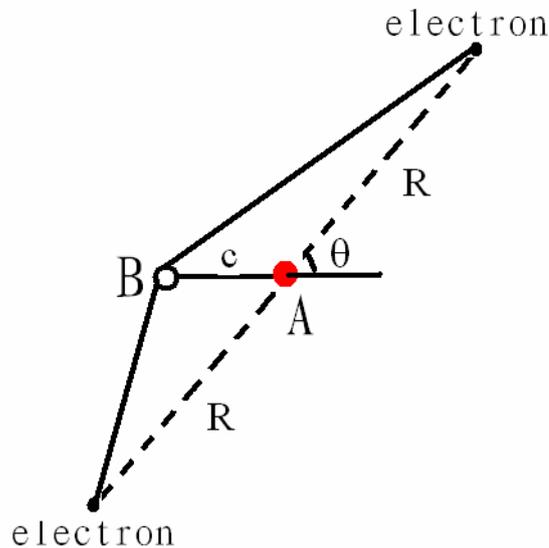

**Fig2. Sketch of the pairing mechanism**

Fig.2 illustrates the additional energy caused by electronic dipole lattice. The average distance 2R between two nearest electrons is equal to the distance between nearest lattice units, so R=0.5 is chosen to stand for the coherence length of pairing electrons without affecting the main result of this paper. When the ion charge transfers from A to B, the resulting energy difference can be directly calculated as:

$$V_e = \frac{1}{\pi}\frac{e^2}{4\pi\varepsilon}\int_0^{\pi}\left(\frac{2}{R} - \frac{1}{\sqrt{(R\cos\theta-c)^2+R^2\sin^2\theta}} - \frac{1}{\sqrt{(R\cos\theta+c)^2+R^2\sin^2\theta}}\right)d\theta \approx -\frac{e^2\cdot c^2}{\pi\varepsilon} \quad (5)$$

This simple model shows that the additional energy between two electrons induced by electronic dipole lattice can be negative, and according to BCS theory, pairing happens in such situation. Since $N_{dipole}$ is less than 0.5, the condition that two electrons paired through an electronic dipole is satisfied.

The energy gap $\Delta$ of cooper pairs is determined by the additional energy induced by electronic dipole lattice, including the energy between paired electrons and that between electronic dipoles. Considering about the coulomb screening, and the interaction between two electronic dipoles is reversely to three times of the distance, only the nearest interaction of electronic dipoles is taken into account, and the energy gap can be approximately written as:

$$\Delta \approx N_{dipole}V_e + E_{dipole}/2 = -\frac{\sqrt{1-P_A^2}}{2}\frac{e^2\cdot c^2}{\pi\varepsilon} + \frac{1-P_A^2}{4}\frac{e^2\cdot c^2}{2\pi\varepsilon} = -\frac{e^2\cdot c^2}{2\pi\varepsilon}\left(\sqrt{1-P_A^2} - \frac{1-P_A^2}{4}\right) \quad (6)$$

From equation (6) one can obviously see that there is a negative energy gap when $P_A<1$, and the energy gap will disappears when $P_A=1$.

It is known that the energy gap of normal metal superconductor comes from the lattice deformation caused by motion of electrons, and in the square lattice model the energy gap is proportional to two times of A-B bond length. Since the bond length between A-B ions is much larger than the lattice deformation of normal metal, as a result, the energy gap induced by electronic dipole lattice can be much larger than that of normal metal superconductor.

In the case of the square lattice with triangle units, the position difference of A-B ions will violate the average charge distribution slightly. The violation can be treated as an electronic dipole lattice, and the negative energy gap induced by electronic dipole lattice is relatively large. It means that, though difficult, the lattice deformation, so the transition temperature of superconductor, can be designed.

In conclusions, the effect of quantum phenomenon on $CuO_2$ plane ---the random transition of electrons between neighbored ions is studied. It is found that effect of this quantum phenomenon will violate the ion charge uniform distribution, and results in attraction between two electrons.   This study also offers a brand new method to find new superconductor material, and maybe helpful to explain the mechanism of cuprate superconductor.

**Acknowledgement**

Thanks professor Li Bai-wen for valuable discussion and the writing.